# Lunar Resources: A Review

Ian A. Crawford, Department of Earth and Planetary Sciences, Birkbeck College, University of London, Malet Street, London, WC1E 7HX.

Email: i.crawford@bbk.ac.uk

**Abstract**

There is growing interest in the possibility that the resource base of the Solar System might in future be used to supplement the economic resources of our own planet. As the Earth's closest celestial neighbour, the Moon is sure to feature prominently in these developments. In this paper I review what is currently known about economically exploitable resources on the Moon, while also stressing the need for continued lunar exploration. I find that, although it is difficult to identify any *single* lunar resource that will be sufficiently valuable to drive a lunar resource extraction industry on its own (notwithstanding claims sometimes made for the $^3$He isotope, which I find to be exaggerated), the Moon nevertheless does possess abundant raw materials that are of potential economic interest. These are relevant to a hierarchy of future applications, beginning with the use of lunar materials to facilitate human activities on the Moon itself, and progressing to the use of lunar resources to underpin a future industrial capability within the Earth-Moon system. In this way, gradually increasing access to lunar resources may help 'bootstrap' a space-based economy from which the world economy, and possibly also the world's environment, will ultimately benefit.

## I Introduction

To-date, all human economic activity has depended on the material and energy resources of a single planet, and it has long been recognized that developments in space exploration could in principle open our closed planetary economy to essentially unlimited external resources of energy and raw materials (e.g. Martin, 1985; Hartmann, 1985; Schultz, 1988; Lewis, 1996; Wingo, 2004; Metzger et al., 2013; a thorough review of the literature as it stood at the end of the 20th century has been given by Hempsell, 1998). Recently, there has been renewed interest in these possibilities, with several private companies established with the stated aim of exploiting extraterrestrial resources (these include companies with the names Astrobiotic Technology, Deep Space Industries, Golden Spike, Moon Express, Planetary Resources, and Shackleton Energy Company; details may be found on their respective websites). As the Earth's closest celestial neighbour, the Moon seems likely to play a major role in these activities.

As a result of multiple remote-sensing missions conducted over the last two decades (see Crawford et al., 2014, for a recent summary), combined with the continued analysis of samples collected by the Apollo and Luna missions 40 years ago (Heiken et al., 1991; Jolliff et al., 2006), we are now able to make a first-order assessment of lunar resource potential. In doing so, it is useful to distinguish between three possible uses for lunar materials: (i) the use of lunar materials to facilitate continued exploration of the Moon itself (an application usually referred to as *In Situ* Resource Utilisation, or ISRU); (ii) the use of lunar resources to facilitate scientific and economic activity in the vicinity of both Earth and Moon (so-called cis-lunar space, including operations in Earth orbit) as well as support for future space activities elsewhere in the Solar System; and (iii) the importation of lunar resources to the Earth's surface where they would contribute directly to the global economy. These three possible applications of lunar resources are not mutually exclusive, although they do represent a hierarchy of increasing difficulty and, therefore, a temporal order of likely implementation.

In this paper I first summarise lunar geology for readers who may be unfamiliar with it (Section II). I then review what is currently known about the resource potential of the Moon, stressing the limitations of present knowledge (Section III), and briefly discuss future exploration activities that will be required if we are to increase our knowledge of lunar resource potential (Section IV). I then discuss how these resources might be economically utilized within the three categories of potential applications identified above (Section V), and the international and legal context within which these developments will take place (Section VI). My conclusions are presented in Section VII.

**II Lunar geology relevant to identifying possible resources**

The lunar surface is divided into two main geological units: the ancient, light-coloured lunar highlands, and the darker, generally circular, lunar mare ('seas') which fill the large impact basins, predominantly on the nearside (Figure. 1). The compositions of these materials are now reasonably well-characterised (see Heiken et al., 1991; Jolliff et al., 2006; Jaumann et al., 2012 for reviews), and this information is summarised in Figures 2 and 3. The lunar highlands are thought to represent the original crust of the Moon, and are composed predominantly of anorthositic rocks (i.e. rocks containing more than 90% plagioclase feldspar). Lunar plagioclase is invariably Ca-rich, and to a first approximation the dominant mineralogy of the highlands is anorthite ($CaAl_2Si_2O_8$), with iron and magnesium-bearing minerals (principally pyroxene and olivine) typically contributing only a few percent by volume. Thus, the lunar highlands are rich in Ca, Al, Si and O, but relatively poor in Mg and Fe (Figure 3).

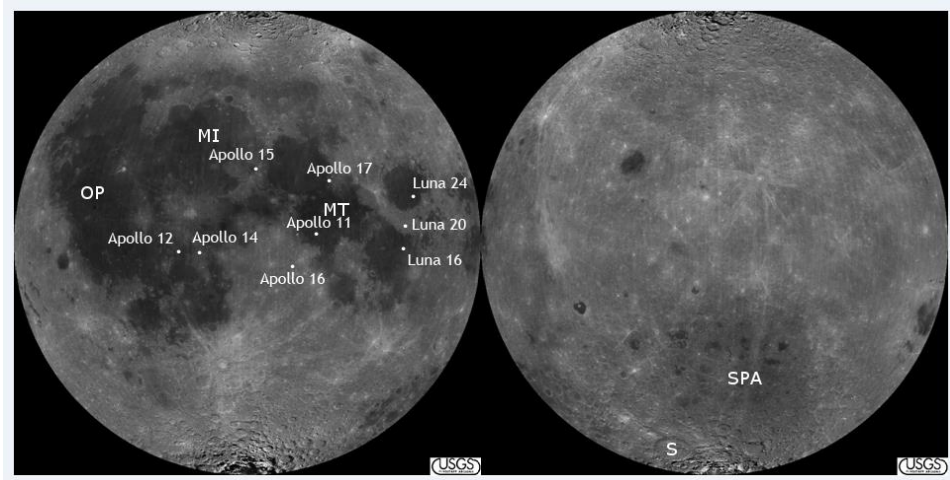

**Figure 1.** Lunar nearside (left) and farside (right) image mosaics, with the landing sites of the six Apollo and three Luna sample return missions indicated. Also marked are locations mentioned in the text: OP: Oceanus Procellarum; MI: Mare Imbrium; MT: Mare Tranquillitatis; SPA: South Pole-Aitkien Basin; S: Schrödinger Basin (image courtesy, USGS/K.H. Joy).

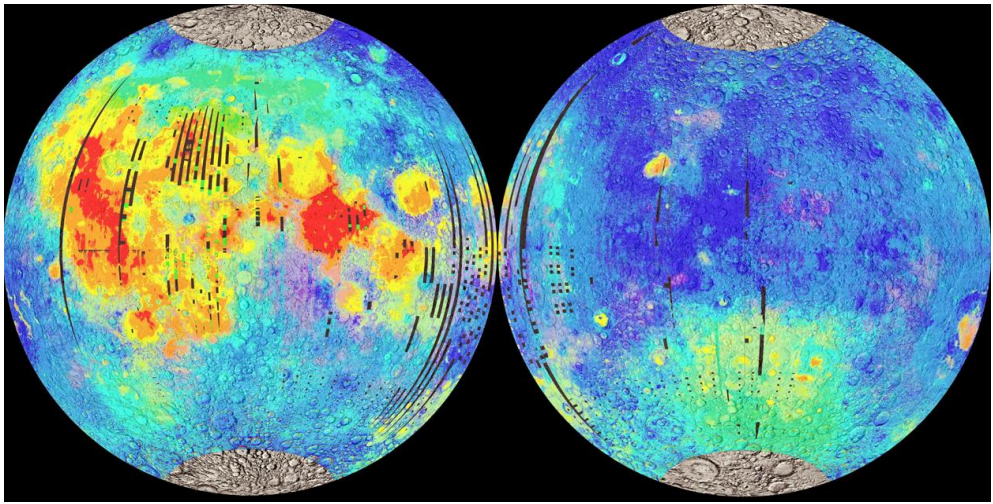

**Figure 2.** Distribution of regolith compositions on the lunar nearside (left) and the farside (right) based on Clementine multi-spectral imaging data. Blue: anorthositic highlands; yellow: low-Ti basalts; red: high-Ti basalts. The large yellow/greenish area in the southern hemisphere of the farside is the South Pole-Aitken Basin, where the colours mostly reflect the more Fe-rich nature of the lower crust exposed by the basin rather than basaltic material (Spudis et al., 2002; courtesy Dr Paul Spudis/LPI).

The lunar maria, on the other hand, are composed of basaltic lava flows. Their mineralogy is dominated by a combination of five minerals (plagioclase [mostly anorthite: $CaAl_2Si_2O_8$], orthopyroxene [$(Mg,Fe)SiO_3$], clinopyroxene [$Ca(Fe,Mg)Si_2O_6$], olivine [$(Mg,Fe)_2SiO_4$], and ilmenite [$FeTiO_3$]), although the proportions of these minerals differ in different lava flows. It follows that the mare basalts are relatively richer in Mg, Fe and Ti, and relatively poorer in Ca and Al (although they are far from lacking in these elements owing to the ubiquitous presence of plagioclase; Figure 3). The principal classification of lunar basalts is based on their Ti content: they are classed as 'low Ti' if their $TiO_2$ abundances are in the range of 1-6% by weight (hereafter wt%), and 'high-Ti' is $TiO_2$ is >6 wt% (Neal and Taylor, 1992). The main Ti-bearing phase in lunar rocks is the mineral ilmenite which, as discussed below, is of particular importance for the retention of some solar wind implanted volatiles and for some proposed ISRU oxygen extraction processes. In this context it is important to note that the ilmenite-bearing high-Ti basalts are mostly confined to two geographically restricted near-side mare regions (i.e. Oceanus Procellarum in the west and Mare Tranquillitatis in the east; see Figure 2), and are therefore not a globally distributed resource.

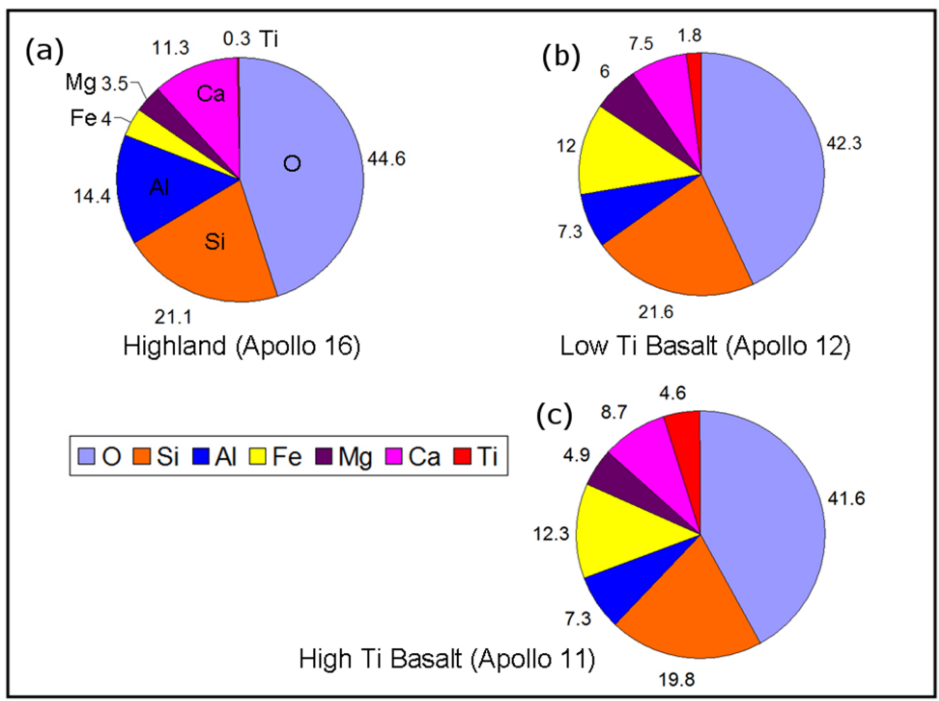

**Figure 3.** Example chemical compositions of lunar soils: (a) lunar highland minerals (Apollo 16); (b) low-Ti basalts (Apollo 12); and (c) high-Ti basalts (Apollo 11). Based on data collated by Stoeser et al. (2010), and reprinted from *Planetary and Space Science,* Vol. 74, Schwandt C, Hamilton JA, Fray DJ and Crawford IA, 'The production of oxygen and metal from lunar regolith' 49-56, Copyright (2012), with permission from Elsevier.

The entire lunar surface is covered in an unconsolidated layer of regolith which has been produced by billions of years of meteorite and micro-meteorite impacts (McKay et al., 1991; Lucey et al., 2006). The regolith is several meters thick in mare regions, and perhaps ten or more meters thick in the older highland areas, and although the uppermost few cm have a powdery consistency (Figure 4) it becomes very compacted (relative density ~ 90%) by a depth of 30 cm (Carrier et al., 1991). The regolith has an average grain size of about 60 µm (McKay et al., 1991), but also contains much larger rock fragments (the sub-cm size fraction is often referred to as the lunar 'soil'). The mineralogical and geochemical composition of the regolith at any given location largely reflects the composition of the underlying rock units, with a small (generally $\leq$ 2%) additional meteoritic component. Understanding the regolith is important for any consideration of lunar resources because it will form the basic feedstock for most processes which may be envisaged for extracting and refining lunar raw materials.

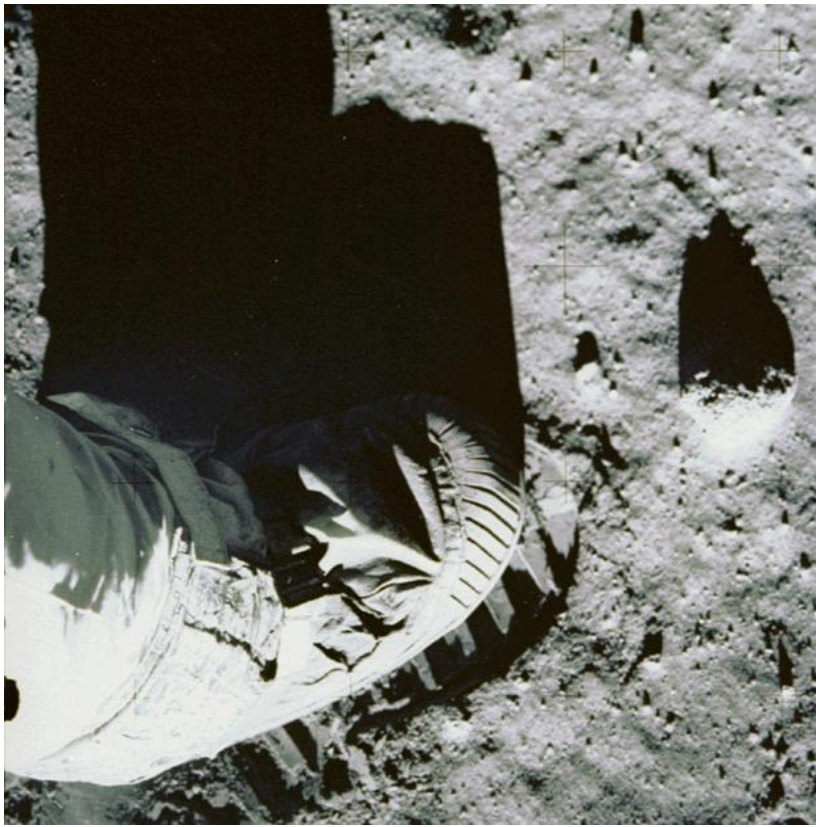

**Figure 4.** Close up view of the lunar regolith with Apollo 11 astronaut Edwin "Buzz" Aldrin's boot for scale, showing consistency of the uppermost surface (NASA image AS11-40-5880).

From a resource perspective, a particularly interesting sub-set of lunar soils are the deposits of volcanic ash, otherwise known as pyroclastic deposits, that appear to have been produced by fire-fountaining volcanic eruptions in some mare areas (e.g. Taylor et al., 1991; Gaddis et al., 2003). Such deposits are identified from orbital remote sensing observations as low albedo ("dark mantle") surficial material surrounding known or presumed volcanic vents (Fig. 5). Over 100 possible lunar pyroclastic deposits have been identified (Gaddis et al., 2003), with twenty having areas greater than 1000 km$^2$. Samples thought to be representative of these materials were collected at the Apollo 15 and 17 landing sites and found to consist of small (tens of µm in diameter) glass, or partially crystallized, beads of essentially basaltic composition. These deposits may have significant resource implications for several reasons: (i) they are relatively enhanced in volatiles compared to most lunar regoliths (e.g. Taylor et al., 1991; Li and Milliken, 2014); (ii) the glass component is more easily broken down for the extraction of oxygen than are crystalline silicates (Taylor and Carrier, 1993); and (iii) they are probably much more homogeneous in size and composition, and less compacted, than the general regolith, which will make their use as a feedstock more straightforward (Hawke et al., 1990).

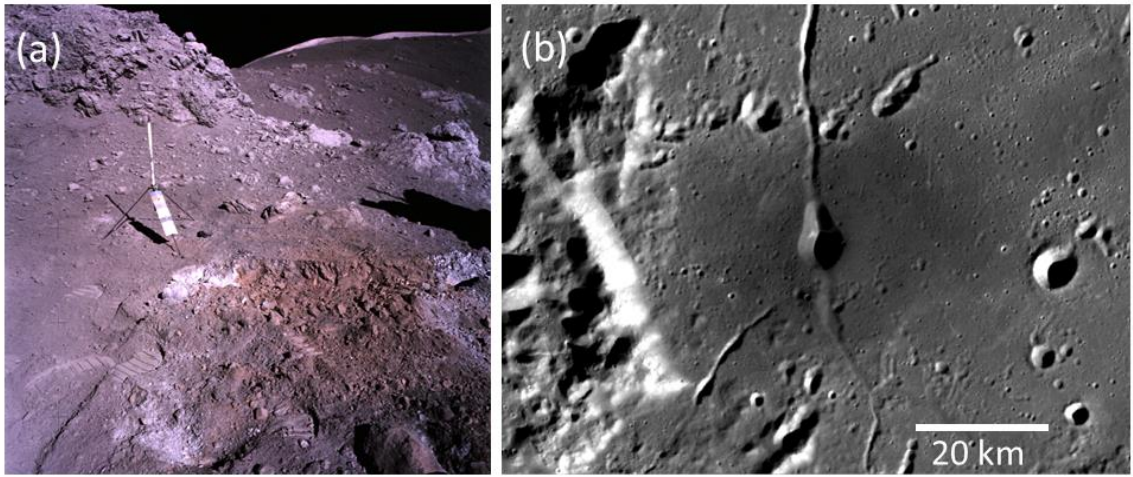

**Figure 5.** Pyroclastic deposits on the Moon. (a) Outcrop of high-Ti orange glass at the Apollo 17 landing site; for scale, the gnomon's feet are about 50 cm apart (NASA image AS17-137-20990). (b) Dark pyroclastic materials surrounding a presumed volcanic vent in the Schrödinger Basin on the lunar farside (see Fig. 1 for location; image obtained by NASA's Lunar Reconnaissance Orbiter Camera, image courtesy NASA/ASU).

With a few exceptions (discussed elsewhere in this paper), the extent to which the lunar crust may contain bodies of locally concentrated 'ores' of economically exploitable materials is still largely unknown. We can be sure that hydrothermal and biological processes, which are responsible for much ore formation on Earth, cannot have operated on the Moon. On the other hand, large quantities of silicate melts, from which economically valuable materials can be concentrated through fractional crystallisation and gravitational settling, have certainly existed in lunar history. Examples include (i) an early 'magma ocean' phase, when most or all of the Moon appears to have been molten (e.g Elkins-Tanton et al., 2011); (ii) subsequent production of magnesium-rich magmas and their intrusion into the lunar crust (e.g. Taylor et al., 1991); (iii) thick sequences of impact melt deposits resulting from basin forming impacts (e.g. Vaughan et al., 2013); (iii) large scale eruption of mare basalts of varying compositions; and (iv) more localised volcanic activity exhibiting an even wider range of magma compositions (including some highly evolved, silica-rich magmas which are otherwise rare on the Moon; e.g. Glotch et al., 2010). These considerations led Papike et al. (1991) to conclude that it is possible that "layered ore deposits similar to or even larger than those of Earth may occur on the Moon."

In addition, although true hydrothermal processes cannot have occurred on the Moon, there is some evidence that volatile-mobilized elements (e.g. F, S, Cl, Zn, Cd, Ag, Au, Pb) have become concentrated by volcanic outgassing responsible for the formation of pyroclastic deposits (e.g. Haskin and Warren, 1991). Moreover, we need to be aware that the unique lunar environment may permit the concentration of economically valuable materials by processes that do not operate on the Earth. This is especially true of the lunar poles, where water and other volatiles may become trapped in permanently shadowed craters (Feldman et al., 1998) and, more speculatively, where processes associated with the electrostatic charging, migration and trapping of metallic dust particles may also occur (e.g. Platts et al., 2013).

### III Potential lunar resources

Haskin et al. (1993) gave a comprehensive geochemical assessment of possible lunar resources based on our knowledge in the early 1990s, and this work has been updated by, among others, Duke et al. (2006), Schrunk et al. (2008), and Anand et al. (2012). Here I summarize the state of current knowledge concerning various different categories of potential resources.

*1 Solar wind implanted volatiles*

Because the Moon has no atmosphere or magnetic field, the solar wind impinges directly onto its surface, and solar wind particles are thereby implanted directly into the lunar regolith. The solar wind dominantly consists of hydrogen and helium nuclei (with a He/H number ratio of about 0.04), with heavier elements comprising less than 0.1 percent by number (e.g. Wurz, 2005). Analyses of samples returned by the Apollo missions have shown that the lunar regolith is efficient in retaining these solar wind implanted ions, which, depending on the age of the surface, have therefore been accumulating for hundreds of millions to billions of years. Studies of the Apollo samples have shown that most of these solar wind implanted volatiles can be degassed from the regolith by heating it to temperatures of between 300 and 900 °C (depending on the element outgassed; Fegley and Swindle, 1993), with 700 °C sufficient to release most of the trapped H and He (Gibson and Johnson 1971; Haskin and Warren, 1991).

Fegley and Swindle (1993) conducted a thorough review of the concentrations of solar wind implanted volatiles in the lunar regolith. The concentrations vary somewhat between the various Apollo and Luna sampling locations, in part owing to the different regolith grain sizes and mineralogies (to which He is especially sensitive). Allowing for this variation, Fegley and Swindle find the average concentrations given in Table 1. There is some evidence, based on the interpretation of orbital neutron spectroscopy (Sinitsyn, 2014), that H concentrations may exceed 100 ppm in some highland areas not represented in the existing sample collection. This is more than twice the average value listed in Table 1, and if solar wind implanted volatiles are to be utilised as a resource then further exploration of such areas may be called for.

It is important to realise that, owing to the continued overturning ("gardening") of the regolith by meteorite impacts, solar wind implanted volatiles are expected to be present throughout the regolith layer, and studies of the Apollo drill cores indicate that their concentrations are approximately constant within the uppermost 2-3 metres (Fegley and Swindle 1993). With this in mind, Table 1 also gives the quantities of these materials expected to be contained within a regolith volume of 1 m$^3$.

Table 1. Average concentrations of solar wind implanted volatiles in the lunar regolith (Fegley and Swindle 1993), where the quoted errors reflect the range (± one standard deviation) of values found at different sampling locations. The corresponding average masses contained within 1 m$^3$ of regolith (assuming a bulk density of 1660 kg m$^{-3}$; Carrier et al., 1991) are also given.

| Volatile | Concentration ppm (µg/g) | Average mass per m$^3$ of regolith (g) |
|---|---|---|
| H | 46 ± 16 | 76 |
| $^3$He | 0.0042 ± 0.0034 | 0.007 |
| $^4$He | 14.0 ± 11.3 | 23 |
| C | 124 ± 45 | 206 |
| N | 81 ± 37 | 135 |
| F | 70 ± 47 | 116 |
| Cl | 30 ± 20 | 50 |

In addition to the volatiles listed in Table 1, lunar soils contain small quantities (typically ≤ 1 µg/g) of the solar wind derived noble gasses Ne and Ar (and much smaller quantities of Kr and Xe). Perhaps more interesting from a resource perspective, they also contain a significant quantity of sulphur (715±216 µg/g; Fegley and Swindle 1993), mostly derived from the mineral troilite (FeS), and this would probably also be released by any process which extracts the other volatile elements.

There is an important caveat to note regarding the global distributions of solar wind implanted volatiles, and that is their unknown concentrations at high latitudes. All the numbers given in Table 1 are based on samples collected at low latitudes on the nearside (Fig. 1). Although high lunar latitudes are exposed to a lower solar wind flux, volatiles may be retained more easily owing to the lower surface temperatures. There is remote sensing evidence (Pieters et al., 2009) that this is the case, although quantifying its importance will require collecting samples, or making *in situ* measurements, at much higher latitudes than has been attempted to-date.

It should be noted that extracting these solar-wind implanted materials will be quite energy intensive. The ambient temperature of the regolith below the uppermost few centimetres is about –20°C (Vaniman et al., 1991), so extracting the bulk of the implanted volatiles will require raising this by about 720°C. The specific heat capacity of lunar regolith is a function of temperature (see discussion by Rumpf et al. 2013), with an average value of about 900 J Kg$^{-1}$ K$^{-1}$. The regolith has a bulk density of approximately 1660 kg m$^{-3}$ (Carrier et al., 1991), so the energy required to raise the temperature of a cubic metre of regolith by 720°C will be of the order of 10$^9$ J. This is approximately the amount of energy which falls on a square metre of the equatorial lunar surface (i.e. a surface perpendicular to the Sun) in nine days. Application of the required energy to the regolith might be achieved by direct concentration of sunlight (e.g. Nakamura and Senior, 2008) or by microwave heating of the soil (Taylor and Meek, 2005). In any case, as many cubic meters of regolith will need to be processed to yield useful quantities of these materials (Table 1), it is clear that a significant processing infrastructure will be required.

As will be discussed in more detail in Section V, once extracted some of these volatile elements clearly have potential uses. For example, solar wind implanted H could be useful as a rocket fuel and as a reducing agent for some of the schemes proposed for extracting oxygen and metal from metal oxides (see below), lunar $^4$He might conceivably be a useful addition to terrestrial helium reserves, and, in the context of long-term human operations on the Moon, a local source of C and N is likely to be required to support lunar agriculture (see discussion by Kozyrovska et al., 2006). However, most interest has focussed on the light isotope of helium, $^3$He, to which we now turn.

*2 Helium-3*

In principle, and given a suitably designed reactor, $^3$He could be fused with deuterium (D) to yield energy through the nuclear reaction:

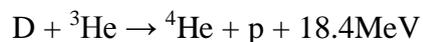

$$D + {}^3He \rightarrow {}^4He + p + 18.4 \text{MeV}$$

where p represents a proton, and 18.4 MeV is the energy released per reaction (1 mega-electron volt, MeV= $1.602\times10^{-13}$ Joules). $^3$He is very rare on Earth whereas D is relatively abundant in sea water (D/H=$1.6\times10^{-4}$), which has given rise to the suggestion that $^3$He might be mined on the Moon and transported to Earth where it could be used in future nuclear fusion reactors employing the above reaction to produce electrical power (Wittenberg et al., 1986; Taylor and Kulcinski, 1999, Schmitt, 2006). Conceivably it might also be useful for nuclear fusion power in space.

The first point to note is that, although it is sometimes claimed that $^3$He is 'abundant' on the Moon (e.g. Tronchetti, 2009), this is actually far from being the case. As shown in Table 1, its average abundance in the regolith is about 4 parts-per-billion (ppb) by mass, with the highest measured concentrations in lunar regolith samples being about 10 ppb (Fegley and Swindle, 1993). Schmitt (2006) has argued that some of the $^3$He originally implanted in lunar samples may have been lost as a result of agitation during collection, transport to Earth, and storage and handling before the measurements were made, and that therefore that the measurements may underestimate the actual concentrations (perhaps as much as 40%). This suggestion is plausible, although proving it will require renewed sampling or *in situ* measurements on the lunar surface. In any case, it does not alter the basic fact that, by any objective standard, the $^3$He isotope is actually very *rare* in lunar soils, having only ppb concentrations.

The second point to note is the non-uniform spatial distribution of lunar $^3$He. Studies of the Apollo samples have demonstrated that both isotopes of solar wind implanted He are preferentially retained in the mineral ilmenite (Fegley and Swindle 1993), from which it follows that $^3$He will be most abundant in regoliths developed on high-Ti mare basalts. In addition, as would be expected, it has been found that the He abundance correlates with measures of the 'maturity' of the lunar regolith (essentially a measure of the length of time to which they have been exposed to the solar wind). As lunar remote sensing measurements can estimate both the Ti concentration and maturity of lunar soils, it is possible to map the estimated $^3$He concentrations in regions that have not yet been sampled (e.g. Fegley and Swindle 1993; Johnson et al., 1999; Fa and Jin, 2007).

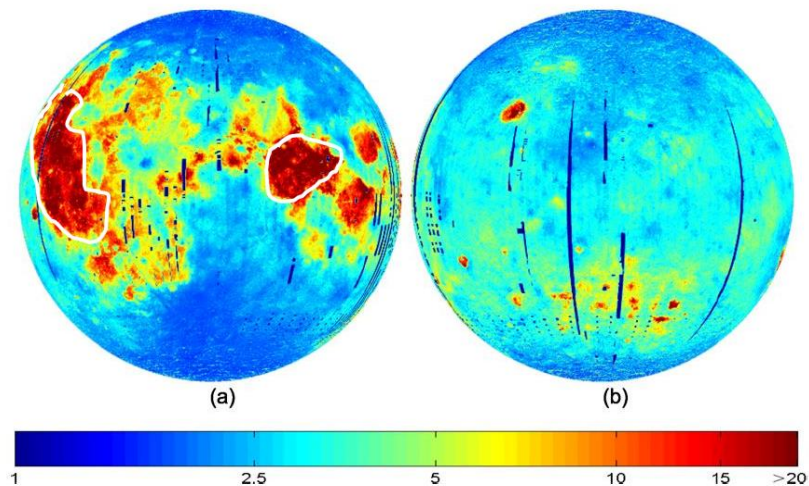

**Figure 6**. Estimated concentration of $^3$He (parts per billion by mass) in the lunar regolith: (a) nearside; (b) farside; the white contours in (a) enclose areas of enhanced $^3$He concentration in Oceanus Procellarum (left) and Mare Tranquillitatis (right; cf. Fig. 1). Reprinted from *Icarus,* Vol. 190, Fa W and Jin Y-Q, 'Quantitative estimation of helium-3 spatial distribution in the lunar regolith layer', 15-23, Copyright (2007), with permission from Elsevier.

Figure 6 shows the results of one such exercise (Fa and Jin, 2007), which clearly shows the expected non-uniform distribution of lunar $^3$He, with the high-Ti mare basalts of Mare Tranquillitatis and Oceanus Procellarum dominating. In these two areas, which collectively cover about 2 million km$^2$, the concentrations of $^3$He may exceed 20 ppb. This is about five times the global average value of 4.2 ppb found by Fegley and Swindle (1993), and is about twice the highest concentration actually measured in a regolith sample to-date. Taking the 20 ppb concentrations estimated by Fa and Jin

(2007) for Mare Tranquillitatis and Oceanus Procellarum, which would clearly be the most economically exploitable deposits, and assuming a regolith thickness of 3 m, yields a total mass of $^3$He for these two regions of about $2\times10^8$ kg .

Finally, it is important to reiterate the point made above regarding the unknown distribution of $^3$He at high latitudes. Although the well-documented correspondence of $^3$He concentration with ilmenite abundance would lead us to expect low polar concentrations, as indicated in Figure 6, the extent to which the lower temperatures of polar regoliths may enhance retention of $^3$He is not currently known.

*3 Water*

The Moon is generally considered to be an anhydrous body, and indeed everything we have learned from the analysis of lunar samples and remote-sensing data indicates that it is highly depleted in water and other volatiles compared to the Earth (e.g. Papike et al., 1991; Haskin and Warren, 1991). That said, in recent years lunar exploration has shown that the Moon's surface environment may contain a number of possibly exploitable water reservoirs (reviewed by Anand, 2010). Clearly these will never be of any economic value to Planet Earth, but they may prove valuable in the context of a future space-based economy (e.g. Spudis and Lavoie, 2011; see discussion in Section V).

Of particular interest are craters near the lunar poles, the floors of which are in permanent shadow owing to the low (1.5°) obliquity of the Moon's rotation axis. It was recognized by Watson et al. (1961) that water ice may be stable in such permanently shadowed 'cold traps', where surface temperatures have now been measured to be below 40 K (Paige et al., 2010). Any water ice found within such cold traps is likely ultimately to be derived from the impacts of comets and/or hydrated meteorites with the lunar surface, although it is also possible that $H_2O$ and/or OH molecules produced by interactions between the solar wind and the regolith at lower latitudes (Ichimura et al., 2012) may migrate to the polar regions and become trapped there.

Indirect evidence for ice in permanently shadowed polar craters was provided by the neutron spectrometer on the Lunar Prospector Spacecraft (Feldman et al., 1998; 2001). This interpretation was supported by the results of NASA's Lunar Crater Observation and Sensing Satellite (LCROSS) mission, which impacted into a permanently shadowed region of the southern polar crater Cabeus in 2009, and measured a water ice concentration of 5.6 ± 2.9 wt% (1 sigma errors) in the target regolith (Colaprete et al., 2010). Supporting observations have come from space and Earth-based radar instruments (e.g. Nozette et al., 2001; Spudis et al., 2013), which may indicate the additional presence of substantial (≥ 2 m thick) deposits of relatively clean ice in some permanently shadowed regions, although the interpretation of these data has been questioned (e.g. Campbell et al., 2006; Starukhina, 2012). Confirmation of the presence of water ice in these regions, and the form the ice takes (e.g. blocks of relatively pure ice, or ice crystals mingled with the regolith) will almost certainly require additional *in situ* measurements.

McGovern et al. (2013) have determined that the cumulative area of permanently shadowed lunar surface is 13,361 km$^2$ in the northern hemisphere and 17,698 km$^2$ in the southern hemisphere, giving a total area of 31,059 km$^2$. As expected, in both hemispheres most areas of permanent shadow occur pole-ward of 80° north or south latitude, but small patches of permanent shadow are identified at latitudes as low as 58° in both hemispheres. The extent to which any or all of these permanently shadowed areas contain water ice and other volatiles is not currently known. Clearly, if they all possessed ice at the level implied by the LCROSS measurements in Cabeus then the total quantity of water could be very high (assuming a concentration of 5.6% by mass and a regolith density of 1660 kg m$^{-3}$, the total mass of water contained within the uppermost meter of permanently shadowed regolith would be 2.9×10$^{12}$ kg (or 2900 million tonnes).

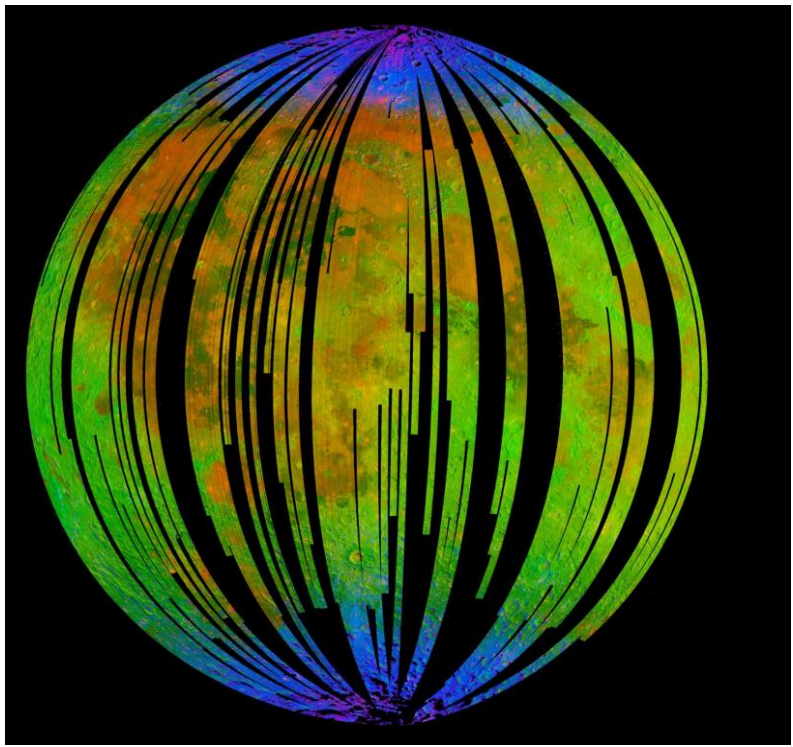

**Figure 7.** False colour image of the lunar nearside based on data obtained by the Moon Mineralogy Mapper (M$^3$) instrument on India's Chandrayaan-1 mission. Blue represents areas where a 2.8 to 3.0 μm absorption band attributed to bound H$_2$O/OH was detected; red represents areas where the 1 μm band of the mineral pyroxene is strong, and picks out the basaltic maria; and green represents albedo. Note that evidence for hydration is restricted to high latitudes, but is much more extensive than areas of permanent shadow at the poles (image courtesy, NASA/ISRO/Brown University/R. N. Clark, USGS/AAAS).

In addition to possible ice in permanently shadowed craters, infra-red remote-sensing observations (Pieters et al. 2009) has found evidence for hydrated minerals (and/or adsorbed water or hydroxyl molecules) covering high latitude, but non-permanently shadowed, areas of the lunar surface (Fig. 7). It is thought that this OH/$H_2O$, which cannot exist as ice in sun-lit areas, is produced by the reduction of iron oxides in the regolith by solar wind-implanted hydrogen (Ichimura et al., 2012), with OH/$H_2O$ being retained in the relatively cold high-latitude regolith. Pieters et al. (2009) estimated that the corresponding water abundance might be as high as 770 ppm in these areas, but this is model-dependent and the actual concentrations could be much lower. Moreover, the extent to which these hydrated materials are purely surficial (and thus probably of negligible value from a resource perspective) or may have been gardened into a substantial volume of the underlying regolith (in which case they may be volumetrically more important) is not currently known and will require future exploratory missions to determine.

Last but not least, we note that there is evidence that lunar pyroclastic deposits have significant levels of hydration. Laboratory measurements of the Apollo 15 and 17 pyroclastic glasses have water contents mostly in the range 5-30 ppm (Saal et al., 2008). These concentrations are very modest from a resource perspective: at 10 ppm 1 $m^3$ of this material would yield only about 20 ml of water. However, the interpretation of recent infra-red remote-sensing observations implies that some pyroclastic deposits may contain as much as 1500 ppm (i.e. 0.15 wt%) of $H_2O$ (Li and Milliken, 2014). If confirmed, such high concentrations may well be economically exploitable because, although an order of magnitude lower than those implied for permanently shadowed craters by the LCROSS results, they occur in much less extreme environments where solar energy will be readily available.

*4 Oxygen*

An indigenous source of lunar water, either in the form of polar ice or in hydrated regoliths and/or pyroclastic deposits, would be the preferred choice as a source of oxygen on the Moon. Nevertheless, it has long been recognized that, if necessary, oxygen could be extracted from anhydrous oxide and silicate minerals in the lunar regolith (see the extensive reviews by Taylor and Carrier, 1993; Schrunk et al., 2008). Indeed, Taylor and Carrier (1993) discuss twenty different possible processes for extracting oxygen from lunar regolith, and produce a short-list of eight which they consider to be the most practical. It does have to be noted that all are quite energy intensive: Taylor and Carrier estimated that their eight shortlisted processes would require between 2-4 megawatt-years of energy (i.e. 6-12×$10^{13}$ J) to produce 1000 tonnes of oxygen. The small scale experiments reported by Li et al. (2012) actually required energy levels about two orders of magnitude higher, although they noted that the efficiency can probably be increased significantly. In any case, these power levels would require a small nuclear reactor or several thousand square metres of solar panels.

Historically, one of the most studied of these processes involves the reduction of the mineral ilmenite, for example:

$$FeTiO_3 + 2H \rightarrow Fe + TiO_2 + H_2O$$

In common with all proposed oxygen extraction schemes, high temperatures (700-1000°C; Taylor and Carrier, 1993; Zhao and Shadman, 1993; Li et al., 2012) are required for this reaction, which in part accounts for the high energies required. Note that this particular route to oxygen produces water in the first instance, and is therefore a possible route to producing water on the Moon if indigenous sources (as described above) are unavailable, provided that a source of hydrogen exists.

Ilmenite is found mostly in high-Ti mare basalts (where it can contribute more than 25% by volume; Papike et al., 1998), so it follows that this process would be most efficiently employed in the same geographical regions discussed above in the context of $^3$He extraction (i.e. Oceanus Procellarum, Mare Tranquillitatis, and additional small patches of high-Ti basalts located elsewhere; Fig. 2). The reaction requires an initial source of hydrogen as a reducing agent (which might itself be extracted from the regolith, as discussed above), but as the last step of the process involves the electrolysis of water to produce oxygen and hydrogen the latter could be recycled. Note that this process also produces Fe metal and rutile ($TiO_2$), which might be further processed to produce Ti and additional oxygen that may also have additional economic value.

Given that ilmenite reduction is only likely to be viable in high-Ti mare regions, it is desirable to identify processes which could extract oxygen from the more ubiquitous anorthositic (i.e. plagioclase-rich) materials found in highland localites. One possibility might be the recently proposed molten salt electrochemistry process described by Schwandt et al. (2012), which, although designed initially with ilmenite reduction in mind, would work equally well on anorthositic feedstock (where it would yield an aluminium-silicon alloy as a by-product rather than titanium; C. Schwandt, personal communication). This would also have the advantage of operating at a relatively low temperature (~900°C, compared to the 1300-1700°C required for processes based on the electrolysis of silicate melts; Taylor and Carrier, 1993). Another possible process would be vapour phase reduction (reviewed by Senior, 1993, and highly ranked in the comparison performed by Taylor and Carrier). In this process the high temperatures required (>2000°C) would be achieved by focussing sunlight in a solar furnace, thereby avoiding problems associated with electrical power conversion and electrolysis, and some conceptual development work on such systems has already been performed (e.g. Nakamura and Senior, 2008; Nakamura et al., 2008).

It is important to bear in mind that, even if the existence of polar ice deposits are confirmed, processes able to extract oxygen from lunar geological materials may still be important in the context of the future economic development of the Moon. For one thing, the difficulties of extracting and processing water ice in permanently shadowed craters, where the temperatures can be lower than 40 K (Paige et al., 2010), may off-set the apparent advantages that ice has as a feedstock. In addition, there are many scientifically and economically interesting localities on the Moon that are located far from the polar regions, so it may be that a local, even if more energy intensive, source of oxygen will be preferable for supporting operations in those localities. Moreover, as we have seen, oxygen extraction from metal oxides also produces possibly useful metals, which using water as a feedstock does not.

*5 Metals*

With the possible exception of some high value trace metals, such as the platinum group metals (PGMs), one would not go into space for a source of metals for use on Earth. Nevertheless, many metals will be essential for constructing industrial and scientific infrastructures on the Moon itself and in cis-lunar space, and we discuss possible sources here.

***Iron and siderophile elements***. Iron (Fe) is abundant in all mare basalts (~14-17 wt%; Papike et al., 1998), but is mostly locked into silicate minerals (i.e. pyroxene and olivine) and the oxide mineral ilmenite. Extracting iron from these minerals will be energy intensive, although, as we have seen, it would be a natural by-product of producing oxygen through ilmenite reduction.

A more readily available source might be native Fe in the regolith, although the concentrations are quite low (~0.5 wt%; Morris, 1980). Native Fe in the regolith has at least three sources (e.g. Morris, 1980): meteoritic iron, iron released from disaggregated bedrock sources, and iron produced by the reduction of iron-oxides in the regolith by solar wind hydrogen. The latter component occurs as nanometre sized blebs (often referred to as 'nano-phase Fe') within impact glass particles ('agglutinates') and is likely to be difficult to extract. Morris (1980) found the average concentration of the other two components to contribute 0.34±0.11 wt%. At least in principle this might be extractable, perhaps through some form of magnetic sieving, but given the very small sizes of the individual Fe particles (generally < 1μm), the practicality of such a process is uncertain. If the practical difficulties could be overcome, a native Fe concentration of ~0.3 wt% is not an entirely negligible amount, corresponding to about 5 kg for 1 $m^3$ of regolith. Moreover, meteoritic iron will be associated with siderophile elements, some of which (e.g. nickel, the PGMs and gold) are valuable owing to their catalytic and/or electrical properties; at concentrations typical of meteoritic Fe, 1$m^3$ of regolith could potentially yield 300g of Ni and 0.5g of PGMs.

Much higher localised concentrations of native Fe, and associated siderophile elements, could potentially be found in the vicinity of any Fe-rich meteorites which partially survived collision with the lunar surface (Haskin et al., 1993; Wingo, 2004). Indeed, Wieczorek et al. (2012) have recently interpreted prominent lunar magnetic anomalies as being due to surviving Fe-rich meteoritic debris. The strongest of these magnetic anomalies, on the northern rim of the farside South Pole-Aitken Basin (Fig. 8), cover an area of approximately 650,000 $km^2$ and, given Wieczorek et al.'s preferred interpretation that they are caused by surviving fragments of an originally ~110 km-sized metallic core of a differentiated asteroid, could potentially indicate the presence of considerable quantities of near-surface metallic Fe (and associated Ni and PGMs).

Smaller, scattered, magnetic anomalies elsewhere on the Moon (Fig. 8) may similarly indicate the presence of Fe-rich material at or near the lunar surface. Only further exploration will determine whether or not this interpretation is correct, and how exploitable such meteoritic debris may be.

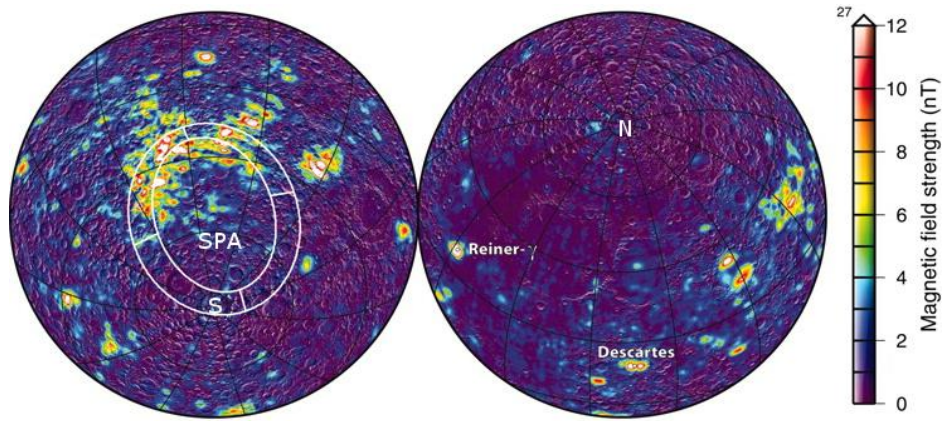

**Figure 8.** Lunar magnetic anomalies found in the Moon's farside southern hemisphere (left) and nearside northern hemisphere (right). The north and south lunar poles are marked 'N' and 'S' respectively; the two white ovals at left denote the inner basin floor and outer structural rim of the South Pole–Aitken basin (SPA) and the outer ellipse has a major axis of 2400 km. Note the occurrence of magnetic anomalies along the northern rim of the basin, and scattered anomalies elsewhere (Reiner-γ and Descartes are two prominent nearside examples). These may represent remains of iron and/or chondritic meteoritic debris in the near sub-surface. From Wieczorek MA, Weiss BP and Stewart ST 'An impactor origin for lunar magnetic anomalies', *Science*, Vol. 335, pp. 1212-1215 (2012), reprinted with permission from the AAAS.

There is an argument (e.g. Lewis and Hutson, 1993; Kargel 1994; Elvis, 2012) that the most easily exploitable extraterrestrial sources of Fe, Ni, and the PGMs will be near-Earth asteroids rather than the Moon. Given the relative ease of access of some of these objects, and their lack of gravity, this reasoning appears sound. Nevertheless, in the context of the future development of the Moon, where an infrastructure may be developed to support human operations for multiple purposes (Section V), access to 'crashed' metallic asteroids (and/or their melted and recrystallized remnants) on the lunar surface may nevertheless prove to be economically valuable.

*Titanium.* Titanium (Ti) is a potentially useful metal for aerospace applications which has a significant concentration (typically in the range 5-8 wt%) in the high-Ti mare basalts of the lunar near-side (Fig. 2). As we have seen, it occurs almost entirely in the mineral ilmenite ($FeTiO_3$), from which it may be extracted by electrochemical processes such as that described by Schwandt et al. (2012). Moreover, unlike the case for Fe discussed above, Ti is a lithophile element and has a very low concentration in metallic asteroids/meteorites (there are only a handful of experimentally measured values in the literature, but most iron meteorites have 'whole rock' Ti concentrations <0.07 wt%, and this figure may reflect the presence of silicate inclusions rather than the concentration in the meteoritic metal itself; Nittler et al., 2004). Ti concentration in chondritic asteroids/meteorites is comparably low (typically <0.09 wt%; Lodders and Fegley, 1998). Therefore, in the context of a future space economy, the Moon may have a significant advantage over asteroids as a source of Ti. The fact that oxygen is also produced as a result of Ti production from ilmenite could make combined Ti/$O_2$ production one of the more economically attractive future industries on the Moon.

*Aluminium.* Aluminium (Al) is another potentially useful metal, with a concentration in lunar highland regoliths (typically 10-18 wt%) that is orders of magnitude higher than occurs in likely asteroidal sources (i.e. ~1 wt% in carbonaceous and ordinary chondtites, and <0.01 wt% in iron meteorites; Lodders and Fegley, 1998; Meyer et al., 2010). It follows that, as for Ti, the Moon may become the preferred source for Al in cis-lunar space. Extraction of Al will require breaking down anorthitic plagioclase ($CaAl_2Si_2O_8$), which is ubiquitous in the lunar highlands, but this will be energy intensive (e.g. via magma electrolysis or carbothermal reduction; Taylor and Carrier 1993; Duke et al., 2006). Alternative, possibly less energy intensive, processes include the fluoridation process proposed by Landis (2007), acid digestion of regolith to produce pure oxides followed by reduction of $Al_2O_3$ (Duke et al., 2006), or a variant of the molten salt electrochemical process described by Schwandt et al. (2012).

## 6. Silicon

Silicon (Si) is abundant in all rocks (about 20 wt% in lunar materials; Fig. 3), and one would not normally think about going into space to obtain it. However, it the context of the future industrialisation of space, it is of potentially enormous importance for the production of arrays of solar cells for the conversion of sunlight into electricity. Some proposed applications (discussed in Section V) will require such large quantities of Si that a space-based source would be desirable. Proposed extraction strategies are similar to those that would be used for extracting Al from lunar materials (i.e. magma electrolysis, carbothermal reduction, fluorination, and/or molten salt electrochemistry; e.g. Taylor and Carrier 1993; Duke et al., 2006; Landis, 2007; Schwandt et al., 2012). The purity requirements for use in semi-conductor devices are very stringent, and may be challenging to achieve in the lunar environment. However, routes to the production of solar cell grade silicon by molten salt electrolysis on Earth are currently under study (e.g. Oishi et al., 2011), and a conceptual outline for solar cell production on the lunar surface has been proposed (e.g. Duke et al., 2001; Ignatiev and Freundlich, 2012).

## 7 Rare earth elements

The rare earth elements (REE; usually defined as the 15 lanthanides, spanning lanthanum to lutetium in the periodic table, plus scandium and yttrium) exhibit a wide range of industrially important optical, electrical, magnetic and catalytic properties (e.g. BGS, 2011; Chakhmouradian and Wall, 2012; Hatch, 2012). It has been known since the return of the Apollo samples that some lunar rocks contain relatively enhanced levels of REE, which, because of corresponding enhancements in phosphorus (P) and potassium (K), have become known in lunar geology as KREEP-rich (Warren and Wasson, 1979). Because uranium and thorium are also concentrated in KREEP-rich materials (see below), their distribution can be mapped by orbiting instruments able to detect gamma-rays emitted by these radioactive elements (e.g. Prettyman et al., 2006). Figure 9 shows a map of lunar surficial Th concentrations as mapped by the Gamma Ray Spectrometer on the Lunar Prospector spacecraft, from which it is evident that KREEP-rich lithologies are mostly associated with Oceanus Procellarum and the Imbrium Basin on the north-western near-side (a region of the Moon that has become known as the Procellarum KREEP Terrain, PKT; Jolliff et al., 2000). In the context of lunar geological evolution, KREEP is thought to have become concentrated in the final stages of mantle crystallization and then exhumed by the impact which excavated the Imbrium Basin.

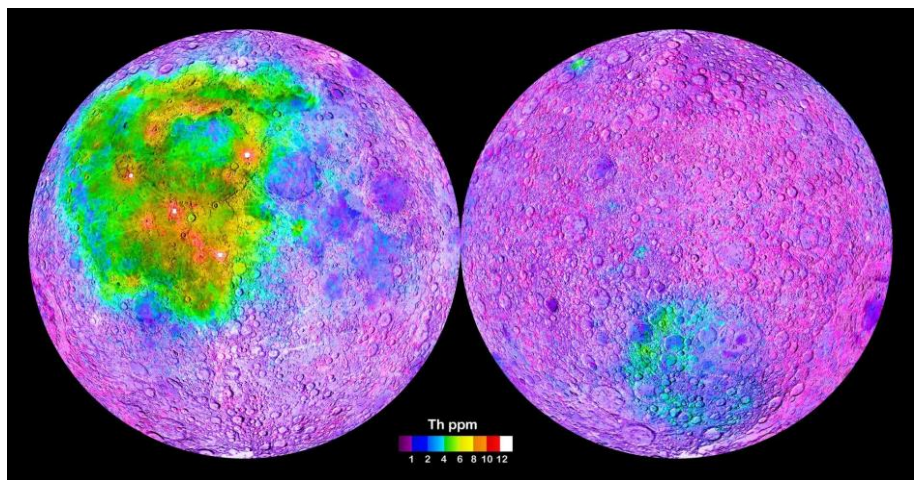

**Figure 9.** Map of lunar surficial Th concentrations on the nearside (left) and farside (right) measured using the Gamma Ray Spectrometer on NASA's Lunar Prospector Spacecraft (Prettyman et al., 2006; image courtesy of NASA).

Warren and Wasson (1979) studied the composition of this mantle reservoir (which they named urKREEP, adopting the German prefix 'ur-' for 'primitive'), and obtained an estimated total REE concentration of about 1200 ppm (0.12 wt%). This total hides considerable variation among individual REE, with yttrium and cerium having concentrations of about 300 ppm each, but europium and lutetium having concentrations of only 3-5 ppm. These REE concentrations are very much at the lower end of the scale for economically exploited REE deposits on Earth, which are generally contain somewhere in the range of 0.1 to 10 wt% REE (BGS, 2011). Moreover, because most lunar crustal rocks, even in the PKT, only contain a component of urKREEP, in most cases the REE concentrations will be lower.

At first sight, this would seem to indicate that the Moon is unlikely to be a useful source of REE. However, it is worth pointing out that some samples from the Apollo 14 and 15 landing sites within the PKT actually have REE abundances several (typically 2 to 4) times higher than the urKREEP values (Haskin and Warren, 1991; Papike et al., 1998). Although still low by the standards of terrestrial ores, it is important to stress that the very localised areas where remote-sensing indicates the highest likely concentrations of REE (essentially the white dots in Fig. 9, where surficial Th has its maximum concentration) have not yet been sampled directly. Moreover, the current spatial resolution of the remote-sensing data is quite coarse (typically several tens of km; Prettyman et al., 2006), and smaller scale outcrops with potentially higher abundances will not be resolved. Future prospecting in these areas may yet identify lunar REE deposits which might be exploitable. However, and in spite of their name, Earth itself has abundant REE deposits (e.g. Hatch, 2012), so it seems unlikely that exporting lunar REE to Earth will be economically viable (unless the *environmental* costs of REE extraction on Earth become prohibitive, a point to which I return in Section V), and lunar REE mining may have to wait until off-Earth markets are developed.

## 8 Thorium and uranium

In principle, uranium (U), either on its own or as a feedstock for the production plutonium (Pu), could prove to be an important element for the development of space-based nuclear power and nuclear propulsion concepts (e.g. Turner, 2005; Dewar, 2007). Conceivably, Th may one day have similar applications, for example in the production of fissile $^{233}$U (Ashley et al., 2012), although such applications have not yet been developed. On the Moon, we expect U and Th to be concentrated in KREEP-rich terrains, and therefore to be spatially correlated with enhancements in the REE discussed above. In lunar materials the abundance of uranium has been found to scale linearly with Th, with a U/Th ratio of approximately 0.27 (Korotev, 1998), so the lunar Th map shown in Fig. 9 also indicates the locations where U enhancements may be expected. This expectation was confirmed when Yamashita et al. (2010) published the first global lunar map of U, obtained with the gamma-ray spectrometer on board the Japanese lunar orbiter Kaguya. This work revealed a maximum surficial U abundance of about 2 ppm in the highest Th-bearing areas of the PKT south of the Imbrium basin.

The Kaguya U map has a spatial resolution of 130 km (Yamashita et al., 2010), and so higher concentrations are likely to exist that are currently unresolved in existing remote-sensing data. For comparison, the urKREEP concentration of U is about 5 ppm (Warren and Wasson, 1979), and some small individual Apollo samples of highly evolved lithologies (i.e. quartz monzodiorites and felsites; Papike et al., 1998) are known with U concentrations in the range 10-20 ppm. However, even the latter concentrations are well below what would currently be considered a low grade U ore on Earth (>100 ppm; WNA, 2012), and only further exploration of high U/Th-bearing terrains on the Moon will determine if higher concentrations exist which might be exploitable in the future.

*9 Other potential lunar resources*

In this section I have summarized some of the more obvious natural resources that the Moon may be able to contribute to human civilisation in the future, but other possibilities may remain to be discovered. Indeed, it is important to realise that every element in the periodic table will exist on the Moon at some level, just as they do on Earth, and as our understanding of lunar geological processes improves we may find other elements that have become locally concentrated. Whether or not these will be economically exploitable will, as always, depend on the value placed on them and on the energy and infrastructure available to support their extraction.

Moreover, although not normally considered as economic resources as such, the natural lunar environment has some properties which might be economically exploitable, at least in the context of lunar development. For example, essentially unprocessed lunar regolith may be required for radiation shielding of human habitats (e.g. Eckart, 1999), and perhaps also for habitat construction if it can be sintered (e.g. Desai et al., 1993; Taylor and Meek 2005), converted into concrete (e.g. Lin, 1985), 3D-printed (e.g. Lim and Anand, 2014), or otherwise turned into useable structural components.

Furthermore, the hard vacuum, and night-time/polar cryogenic temperatures of the lunar environment may positively facilitate some industrial processes, such as vacuum vapour deposition, element concentration via ion sputtering (essentially industrial-scale mass-spectroscopy), and the manufacture of ultra-pure materials (Schrunk et al., 2008). Finally, the unadulterated sunlight (available for >90% of a year at some polar localities; e.g. Speyerer and Robinson, 2013) will provide essentially unlimited supplies of solar energy. Indeed, solar energy has the potential to be a significant lunar export, as it could be collected by solar panels on the lunar surface and beamed to Earth and other locations in cis-lunar space (e.g. Criswell, 1998; Criswell and Harris, 2009).

**IV The need for continued exploration**

Any discussion of lunar resource potential needs to recognize that our knowledge of the Moon's geology is still very incomplete. All of the discussion to-date has been based on the study of samples returned from just nine localities (all at low latitudes and on the nearside; Figure 1), and on the interpretation of orbital remote-sensing data which only probe the uppermost few meters (and often only the uppermost few microns) of the lunar surface. It is entirely possible that further exploration will reveal additional concentrations of economically useful materials on the Moon, and indeed the history of geological exploration on our own planet would lead us to expect this.

It follows that, in order to properly assess the extent to which the Moon may host economically useful materials, an extensive future programme of exploration will be required. Key objectives of such a programme would include: (i) confirmation of the presence of water ice and other volatiles in permanently shadowed polar regions, and assessment of the total polar water inventory; (ii) assessment of the extent to which high-latitude, but non-permanently shadowed, regoliths contain hydrated materials, and especially the depth to which such hydration extends below the surface; (iii) assessment of the volatile inventory of lunar pyroclastic deposits; (iv) *in situ* investigations of impact melt sheets, lava lakes, and magmatic intrusions to assess the extent to which fractional crystallization may have enhanced concentrations of economically valuable minerals; (v) *in situ* investigations of lunar magnetic anomalies to determine whether or not they indicate the locations of exploitable concentrations of meteoritic iron and associated siderophile elements; and (vi) *in situ* investigations of high concentrations of KREEP-rich areas to determine whether or not they contain exploitable concentrations of rare earth elements, uranium and thorium.

In addition to further exploration, pilot studies will be required to determine the practicality and efficiency of extracting potential resources we already know about. Such pilot studies, designed to operate on several m$^3$ of regolith, could include (i) *in situ* heating of regolith to ~700°C to demonstrate the extraction and separation of solar wind implanted volatiles; (ii) demonstration of *in situ* oxide reduction and/or electrochemical processes for the extraction of oxygen, and production of metal, from the regolith; (iii) magnetic sieving of regolith to assess the extent to which metallic iron particles might be extracted; and (iv) assuming that polar ice is confirmed, demonstration of in situ extraction of water and subsequent electrolysis to yield hydrogen and oxygen.

Such an extensive programme of investigations will doubtless take many decades, although a start can be made by new robotic lunar missions that are already in the planning stages (summarized by Crawford and Joy, 2014). Within the next decade these space agency-led activities are likely to be supplemented by an increasing number of private ventures, such as those stimulated by the Google X-Prize (Hall et al., 2013; Bowler, 2014), many of which have an explicit focus on resource identification. In the longer term, it seems clear that exploration on this scale would be greatly facilitated by renewed human operations on the lunar surface (e.g. Spudis, 1996, 2001; Crawford, 2003).

**V Applications for lunar resources**

*1 In Situ Resource Utilisation (ISRU) on the Moon*

There are considerable scientific reasons for wanting to establish a scientific infrastructure on the lunar surface, at least comparable to that which currently exists in Antarctica (e.g. Taylor et al., 1985; Spudis, 1996; Eckart, 1999; Crawford et al., 2003, 2012; Ehrenfreund et al., 2012; McKay, 2013; Crawford and Joy, 2014). In addition, as discussed below, a lunar surface infrastructure may become necessary to support economic activities in cis-lunar space. Moreover, far-fetched as it may seem at present, in the context of growing interest in the possibilities of space tourism the coming decades may witness an emerging market for the transport of fare-paying passengers to the lunar surface (e.g. Benaroya, 2010; Economist, 2012; Moskowitz, 2013). Whatever the reasons for establishing a human infrastructure on the lunar surface, there is no doubt that it would benefit from ISRU, simply because any resources obtained from the Moon itself will not need to be hauled up from the Earth (Figure 10).

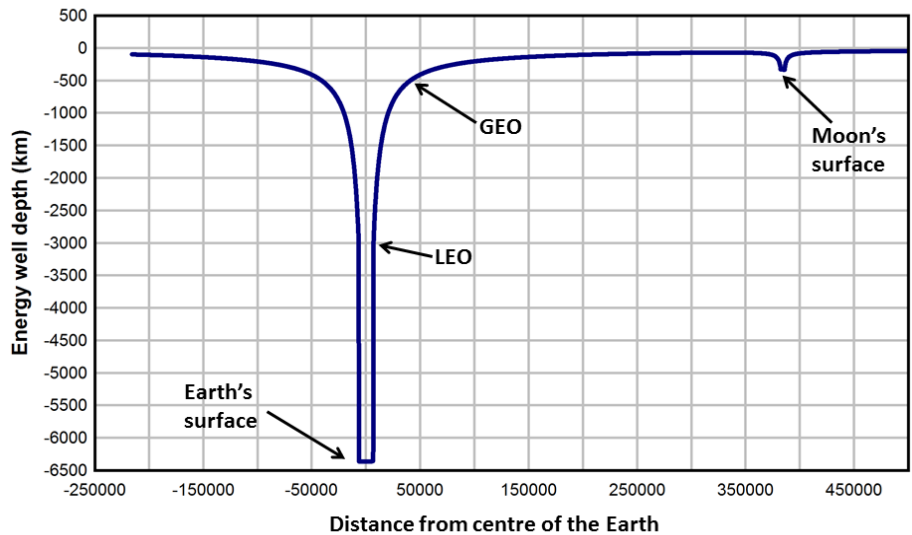

**Figure 10.** Orbital and potential energies in the Earth-Moon system. The vertical axis shows the 'well depth' expressed in units of km [i.e. reaching a stable orbit at a given distance from the Earth would require the same energy as would be expended by climbing against a constant one Earth gravity (9.8 ms$^{-2}$) for the corresponding vertical distance; to convert to energy units (i.e. Joules per kg of lifted mass) multiply by 9800]. For example, escaping the effective gravitational influence of the Earth (the zero level in the diagram) requires the same energy as climbing a 6400 km high mountain; the corresponding value for the Moon is only 290 km (i.e. it requires 22 times less energy to escape from the Moon as from the Earth). Note that all locations in cis-lunar space, including low Earth orbit (LEO) and geostationary orbit (GEO), require far less energy to access from the surface of the Moon than from the surface of the Earth.

Probably the most important, at least in the early stages of operations, would be water for drinking (given suitable purification), personal hygiene, and, in the fullness of time, agriculture. Water would also be an important local source of oxygen for life-support, and water-derived oxygen and hydrogen would also help reduce transport costs if used as rocket oxidiser/fuel. Other volatiles such as C and N, whether obtained from polar cold traps or released from the regolith by heating, will also be important, especially if lunar agriculture is contemplated. Local sources of building materials will also be desirable at an early stage. Initially small-scale scientific outposts (and/or hotels) are unlikely to require the local refinement of metals (although, as noted above, some oxide-reduction schemes will produce these in any case as a by-product of oxygen production). In the longer term, if a substantial lunar infrastructure is developed to support wider economic activities in cis-lunar space, then local sources of metals, semi-conductors (for solar arrays), and even uranium (for nuclear power) would become desirable in addition to indigenous volatiles and building materials.

## 2 Use of lunar resources in cis-lunar space

It has long been recognized (e.g. Ehricke, 1985; Spudis, 1996; Koelle, 1992; Eckart, 1999; Spudis and Lavoie, 2011; Metzger et al., 2013) that lunar resources will be valuable, and perhaps essential, in the future economic development of near-Earth space. This is because, while establishing resource extraction industries on the Moon will be a significant investment, the energy required to transport materials from the lunar surface to any location in Earth or lunar orbit, or indeed anywhere else in the inner Solar System, will be much less than the cost of lifting these materials out of Earth's gravity (Fig. 10).

Our global civilisation is already highly dependent on Earth-orbiting satellites for communications, navigation, weather forecasting and resource management, and this reliance is only likely to increase. The currently high costs of these activities are largely dictated by high launch costs (~ US$ 5000/kg even for the most economic launch vehicles currently available; see discussion by Spudis and Lavoie, 2011), and by the fact that failed satellites, or satellites that have exhausted station-keeping propellants or cryogens, cannot be repaired or replenished in orbit. The availability of resources obtained from the much shallower potential well of the Moon (Fig. 10) would help mitigate these obstacles to further economic development in Earth orbit. Near-term lunar exports to a cis-lunar infrastructure could include the supply of rocket fuel/oxidiser (such as hydrogen and oxygen; especially oxygen which dominates the mass budget for liquid hydrogen/oxygen propulsion), and simple structural components. Later, as the lunar industrial infrastructure becomes more mature, the Moon may be able to provide more sophisticated products to Earth-orbiting facilities, such as Ti and Al alloys, silicon-based solar cells, and uranium (or plutonium derived from it) for nuclear power/propulsion systems.

One particular area where significant future expansion of economic activity in cis-lunar space may occur would be the development of solar power satellites (SPS; Glaser, 1977) in geostationary orbit (GEO). By efficiently capturing sunlight, converting it into microwave energy, and transmitting it to ground stations on Earth, SPS have the potential to become an important component in Earth's future energy mix (Glaser et al., 1998; Flournoy, 2012). However, as noted by Maryniak and O'Neill (1998), the cost of transporting all the materials required to establish a global network of SPS from Earth's surface to geostationary orbit (GEO) may be prohibitive, and a SPS programme would be exactly the kind of large-scale cis-lunar industrial activity that would benefit from the use of lunar resources (again note the small energy differential which exists between the surface of the Moon and GEO compared to that between the surface of the Earth and GEO; Fig. 10).

*3 Use of lunar resources on Earth*

Although a developing cis-lunar economy will necessarily yield wider economic benefits to the Earth as well, the opportunities for lunar resources to make a *direct* contribution to the world economy in the foreseeable future are quite limited. This is because our planet contains the same basic mix of elements as the Moon and the rest of the Solar System, many of them in higher localised concentrations (i.e. ores) and, unlike the Moon, already has a well-developed infrastructure for extracting and refining raw materials. There appear to be three possible exceptions to this general observation, which I discuss below: (i) rare elements for which the market value may render lunar sources economic; (ii) lunar-derived energy sources; and (iii) materials for which the environmental cost of terrestrial mining may make lunar sources more attractive.

***Lunar sources of rare elements.*** Much thought has been given to identifying materials for which their present high market value may render the exploitation of extraterrestrial sources economic (e.g. Lewis, 1996). The PGMs usually feature prominently in these discussions, with near-Earth asteroids (NEAs) identified as possible sources (Lewis and Hutson, 1993; Kargel 1994; Elvis, 2012). However, the possibility of finding 'crashed' iron meteorites on the Moon (e.g. Wingo, 2004; Wieczorek et al., 2012; see the discussion in Section III(5) above) means that lunar sources should not be excluded from consideration. Moreover, although the Moon has a much higher gravity well than do NEAs, it has at least two advantages which may off-set this relative inconvenience. Firstly, it is much closer than any NEA (many NEAs, while they may transiently pass close to Earth, spend most of their orbits at much greater distances and have long synodic periods). Secondly, as discussed above, in the coming decades a local infrastructure may develop on the Moon to support a mix of scientific, industrial and tourism activities, and this could provide local support for mining operations of a kind unlikely to be available at NEAs. Indeed, the presence of such a local infrastructure may reduce the cost and risks of extracting these materials on the Moon to the point that the Moon becomes economically more attractive than NEAs.

One potentially important future application of lunar PGMs may prove to be as catalysts for a future fuel-cell-enabled hydrogen economy on Earth (e.g. Wingo, 2004), where the high cost of these materials is currently a limiting factor (Debe, 2012). Ultimately, whether or not lunar (or indeed any extraterrestrial) sources of PGMs, or other rare and expensive materials, will prove economic in the long-run will depend on whether the prices that these materials can fetch on Earth will remain sufficiently high in the face of a significant increase in supply. Addressing that topic would require an economic analysis beyond the scope of the present paper.

***Lunar sources of energy.*** Most discussion of lunar energy sources for planet Earth have focussed on the possible use of solar wind-derived $^3$He in terrestrial nuclear fusion reactors (Wittenberg et al., 1986; Kulcinski et al., 1988; Taylor and Kulcinski, 1999; Schmitt, 2006; see Section III(2) above). However, there are a number of serious problems with this scenario. Firstly, nuclear fusion has not yet been demonstrated to be a viable source of energy on Earth, and current large-scale experiments (such as the international ITER facility due to become operational in 2028; Butler, 2013) are based on deuterium-tritium (D – $^3$H) fusion rather than D – $^3$He. An oft-stated advantage of $^3$He fusion is the lack of neutrons produced by the primary reaction (which would induce radioactivity in the reactor structure), but as some level of D-D fusion cannot be prevented in a D – $^3$He plasma (Close, 2007) this claimed advantage is often exaggerated (it would apply to pure $^3$He – $^3$He fusion, but that reaction is requires much higher temperatures to initiate).

Even assuming that D – $^3$He fusion *is* found to be technically practical as a power source, the low concentration of $^3$He in the lunar regolith (Section III(2)) implies that mining very large areas would be required. In order to estimate the quantity of regolith that would need to be processed we need to estimate the end-to-end efficiency of converting lunar $^3$He rest-mass energy into electricity on Earth. Conventional power stations generally have efficiencies of about 30%, but advocates of $^3$He fusion estimate efficiencies of 60 – 70% for this process based on the (hypothetical) direct electromagnetic conversion of the fusion product energy into electricity (Kulcinski et al., 1988; Schmitt, 2006). However, this figure ignores the efficiency of extracting $^3$He from the regolith; as discussed in Section III(1), this will require heating large volumes to about 700°C (which would itself consume about 5% of the energy obtained from the fusion of the released $^3$He unless efficient ways of recycling heat from one regolith batch to another can be implemented), and complete efficiency in the release, collection, and purification of $^3$He can hardly be expected (~80% may be reasonable based on the figures given by Kulcinski et al., 1988). Furthermore, transport to Earth will also require energy, as will the extraction, processing and transport of D obtained from sea water to act as a reaction partner (which has been entirely neglected here). Bearing all this in mind, it is hard to see how the end-to-end efficiency could exceed 50%, and may be much less.

Annual world electricity consumption is predicted to rise to $1.4 \times 10^{20}$ J (40,000 TW-hours) by 2040 (US Energy Information Administration, 2014), and if lunar $^3$He were required to produce, say, 10% of this then, for an assumed overall efficiency of 50%, about 500 km$^2$ of high-concentration (i.e. 20 ppb $^3$He) regolith would need to be processed every year down to a depth of 3 meters. Note that, owing to the ubiquitous presence of craters and other obstacles on the lunar surface, only a fraction of a given surface area (estimated at 50% by Schmitt, 2006) will actually be amenable to processing, and allowing for this the highest concentration deposits of this non-renewable material (Fig. 6) would last for about 2000 years. Producing *all* of Earth's anticipated mid-21$^{st}$ century electrical energy requirements from lunar $^3$He would require processing 5000 km$^2$ per year, which may be impractical, and then the accessible reserves would hold out for only about 200 years.

It therefore seems that, at best, lunar $^3$He could only ever make a relatively small contribution to Earth's total long-term energy needs. Given the uncertainty as to whether the world economy will ever come to rely on nuclear fusion as a power source, and the practicality of $^3$He fusion in particular, not to mention the scale of the investment that will be required to extract lunar $^3$He and its non-renewable nature, it appears at least premature to identify lunar $^3$He as a solution to the world's energy needs. That said, assuming the question of practicality can be resolved, the high energy density of $^3$He as a fuel (~5.8×10$^{14}$ J kg$^{-1}$ for D – $^3$He fusion), and its corresponding potential value (several million US dollars per kg at current wholesale energy prices, with the exact value depending on the energy sector with which the comparison is made; US Energy Information Administration, 2014), may nevertheless result in the identification of economically viable, if relatively small-scale, future applications. Possibilities might include space-based fusion power systems, and mobile low-radioactive fusion reactors for use on Earth (e.g. Schmitt, 2006). Moreover, there is currently a shortage of $^3$He on Earth for non-fusion applications (e.g. Kramer, 2010), and lunar $^3$He could possibly help satisfy this demand. As $^3$He will automatically be released as a by-product of extracting other, more abundant, solar wind-implanted volatiles from the lunar regolith (Section III(1)), it is worth keeping an open mind to such possibilities.

As far as the long-term energy requirements of planet Earth are concerned, it would appear more logical instead to invest in developing genuinely inexhaustible energy sources on Earth itself. Possibilities include some combination of D-T and D-D fusion, and/or solar, tidal or geothermal power, supplemented if necessary by space-based solar power systems. As noted above, lunar resources could contribute to the latter either by supporting the construction of geostationary solar power satellites (Maryniak and O'Neill, 1998), or through the construction of solar arrays on the lunar surface and beaming the energy to Earth (Criswell, 1998; Criswell and Harris, 2009). Just to put this latter point in perspective, covering a given area of the near-equatorial lunar surface with solar panels would yield as much electrical energy in 7 years (3×10$^{10}$ J m$^{-2}$, assuming a conservative 10% conversion efficiency) as could be extracted from all the $^3$He contained in the 3m of regolith below it (assuming 20 ppb $^3$He and an optimistic 50% overall efficiency). If the conversion efficiencies were the same (say 20% each, which in practice might not be unreasonable) then solar energy would out produce $^3$He in just 1.4 years. But whereas, once extracted, the $^3$He would be gone forever, the solar power would be continuous. Looked it from this perspective, it would be far more efficient to beam lunar solar energy to Earth than it would be to extract, and physically transport to Earth, $^3$He for hypothetical fusion reactors which have not yet been shown to be practical.

**Environmental benefits of extraterrestrial resources.** The environmental costs of resource extraction on Earth are becoming of increasing concern, and these pressures will only increase as world population rises towards an estimated ten billion people by mid-century (e.g. Tollefson, 2011), all of whom will aspire to increased standards of living. Indeed, if the economic value (to say nothing of the wider societal value) of damaged ecosystems were included in the extraction costs then many sources of raw materials on which the world economy depends would be much more expensive than they are, and some might not be economic at all (e.g. Costanza et al., 1997).

It follows that, when the true value of Earth's natural environment is properly taken into account, the relative cost of exploiting extraterrestrial raw materials may not be as high as first appears. Moreover, preservation of Earth's rich natural habitats is an important social and ethical imperative irrespective of merely economic considerations. Thus, the time may come when the higher costs of extracting lunar (and other extraterrestrial) resources may be off-set by increasingly unacceptable environmental costs of continuing to extract and refine them on Earth. Possible lunar resources to which this argument may come to apply might include the REE (Section III(7)), for which the world economy is certain to demand increasing supply over the coming decades owing to their many technological applications (e.g. Hatch, 2012) but which can be environmentally damaging to extract (BGS, 2011; Chakhmouradian and Wall, 2012). Indeed, as stressed by Bernasconi and Bernasconi (2004), in the long term it is only by accessing space resources that the human burden on Earth's biosphere can be lightened and eventually reduced. However, regardless of the possible environmental benefits, the *option* of shifting resource extraction from Earth to extraterrestrial sources will only be possible in the context of a developing space-based economy, which (as discussed above) is itself likely to depend on lunar resources.

**VI The international and legal context**

In addition to the obvious technical challenges, the development of lunar (and other extraterrestrial) resources will require the establishment of an international legal regime which encourages large-scale investment in prospecting and extraction activities, while at the same time ensuring that space does not become a possible flashpoint for international conflict. On the latter point, Hartmann (1985) has wisely cautioned that "space exploration and development should be done in such a way so as to reduce, not aggravate, tensions on Earth", and this will require the development of an appropriate, internationally supported, legal framework.

Currently space activities are governed by a small set of international treaties negotiated under United Nations auspices (UN, 1984). The most important of these, and the foundation on which the international legal regime for outer space is built, is the 1967 *Treaty on Principles Governing the Activities of States in the Exploration and Use of Outer Space, including the Moon and Other Celestial Bodies*, more commonly known as the Outer Space Treaty (OST). The OST now has 102 States Parties, including all nation-states having a spacefaring capability. As its name suggests, the OST was intended to establish a number of important 'principles' for the activities of nation-states in space. These include the concept that space should be considered "the province of all mankind" (Article I), that outer space is free for the 'exploration and use' by all states (Article I), that the Moon (and other celestial bodies) cannot be appropriated (by claim of sovereignty or otherwise) by nation-states (Article II), and that international law, including the UN Charter, applies to outer space (Article III). This has proved to be an excellent foundation for international space law, but it is clearly a product of its time in that it makes no explicit reference to the exploitation of space resources, or to other commercial space activities that were not envisaged in 1967.

Most importantly in the current context, although it is a clear principle of the OST that nation-states cannot 'appropriate' the Moon as a celestial body, it seems certain that private companies will not invest in the exploitation of space resources unless they can be guaranteed a return on their investment. This in turn implies that they must have legal entitlement to at least the resources they extract (e.g. Pop, 2012). This however then has to be squared with the principle established by the OST (Article I) that "the exploration and use of outer space …… shall be carried out for the benefit and in the interests of all countries," which implies some kind of international regulatory framework. Currently there is ambiguity in the interpretation of the OST in these respects (Tronchetti, 2009; Viikari, 2012; Pop, 2012; Simberg, 2012), and this ambiguity is unhelpful as it acts to inhibit investment. Simberg (2012) explains the problem well when he writes "[w]hile the technology is a challenge, one of the biggest business uncertainties these new [commercial space] companies face is the legal status of any output from their off-world mining operations, and the corresponding ability to raise the funds needed to make them successful."

The one serious attempt to set up a legal regime which went beyond the OST in this respect (namely the 1979 *Agreement Governing the Activities of States on the Moon and Other Celestial Bodies*, otherwise known as the 'Moon Treaty') has failed to gather the support of any spacefaring nation because it was perceived as being overly restrictive on commercial activities (Tronchetti, 2009; Viikari, 2012; Simberg, 2012; Delgado-López, 2014). Although there are reasons to believe that some of the opposition to the Moon Treaty was based on an exaggerated view of these alleged restrictions (e.g. Goldman, 1985), it is clear that a new approach is now required. This is not the place to review all the suggestions that may be found in the space policy literature, but there is a spectrum of possibilities from extreme laissez-faire positions that would repudiate not only the Moon Treaty but possibly even the OST itself, to more internationalist approaches that would retain the spirit of the Moon Treaty while attempting to make its basic approach more compatible with commercial activities in space.

An example of the former position is given by Simberg (2012), who advocates that the US (and by inference other nation-states) unilaterally establish legislation recognizing extraterrestrial land claims (albeit under carefully controlled conditions, including actual occupation of a site on the Moon and/or asteroid by the claimant) as a means of encouraging private investment. However, it is difficult to see how such unilateral action would be acceptable to the international community (or consistent with the OST; see, for example, Tronchetti, 2014), and it would run the risk of violating Hartmann's (1985) injunction that space development should not increase political tensions on Earth. An intermediate position, recognizing the need for international cooperation, but also the difficulty of negotiating a new treaty, calls for the development of international

'norms of behaviour' in space, which would recognize the interests of commercial companies while remaining within the spirit of current international law (e.g. Tronchetti, 2014; Delgado-López, 2014). Finally, at the other end of the spectrum, Tronchetti (2009) has put forward a detailed suggestion for the creation of an 'International Space Authority' to manage the exploitation of lunar (and other extraterrestrial) resources, while also incentivising private investment and protecting sites of special scientific importance.

Whatever legal framework is eventually adopted, if humanity is to reap the potential rewards of utilising the material and energy resources of the Solar System, it is imperative that these outstanding legal and political issues be addressed in a timely manner.

**VII Conclusions**

From the discussion above, it is clear that the Moon does possess abundant raw materials that are of potential economic value for future human activities, and especially for future activities in space. It is true that, on the basis of current knowledge, it is difficult to identify any *single* lunar resource that is likely to be sufficiently valuable to drive a lunar resource extraction industry which has near-term profit as an objective. In particular, I have argued that claims for lunar $^3$He in this respect, through its alleged ability to make a major contribution to Earth's future energy needs, have been exaggerated. Rather, I have identified a hierarchy of applications for lunar resources, beginning with the use of lunar materials to facilitate human activities on the Moon itself (including scientific exploration, resource prospecting, and, perhaps, tourism), progressing to the use of lunar resources to underpin a growing industrial capability in cis-lunar space. In this way, gradually increasing access to lunar resources may help 'bootstrap' a space-based economy from modest beginnings on the lunar surface until much of the inner Solar System lies within the sphere of human economic activity (e.g. Wingo, 2004; Spudis and Lavoie 2011; Metzger et al. 2013). In the fullness of time, this has the potential to yield significant economic benefits to human society on Earth by opening the world economy to external sources of energy and raw materials.

Of course, the Moon is not the only source of extraterrestrial raw materials which could contribute to the development of such a space-based economy, and NEAs are often discussed in this context. However, while resources from NEAs will doubtless come to play an important role, the Moon is likely to remain of central importance because of its constant proximity, its probable concentrations of accessible volatiles (especially water ice at the poles), and the fact that diverse lunar geological processes have concentrated many economically important materials (e.g. Ti, Al, the REEs, and U) in crustal reservoirs to levels which far exceed those found in most known asteroid classes. The Moon also has abundant energy, in the form of sunlight, to extract and process these materials *in situ*. Moreover, although it has long been considered that NEAs would be the preferred extraterrestrial sources for iron, nickel and the PGMs (e.g. Kargel, 1994; Elvis, 2012), the possibility that the Moon may retain partially intact iron and/or chondritic meteorites (Wingo, 2004; Wieczorek et al., 2012) means that it may become a useful source for these materials as well. This is likely to be especially true if a local infrastructure is developed to support the utilisation of other lunar materials. Indeed, it is the fact that the lunar surface has the potential to support a growing scientific and industrial infrastructure, in a way that asteroidal surfaces do not, which is likely to make the Moon the linchpin of humanity's future utilisation of the Solar System.

Finally, it is necessary to reiterate that our knowledge of lunar geological evolution, and therefore of the extent to which the Moon may offer economically useful resources in addition to those identified above, is still very incomplete. Gaining a more complete picture of the Moon's economic potential will require a much more ambitious programme of lunar exploration than has been conducted to-date, and this will probably require the return of human explorers to the lunar surface. Fortunately, such a programme is under active international discussion. In 2013 the world's space agencies came together to develop the Global Exploration Roadmap (ISECG, 2013), which outlines possible international contributions to human and robotic missions to the Moon, NEAs and, eventually, Mars. Implementation of this Roadmap will provide many opportunities for the detailed scientific investigation of the Moon which, in addition to numerous scientific benefits (e.g. Crawford et al., 2012, and references cited therein), will also result in a much clearer picture of lunar resource potential.

## Acknowledgements

I thank my colleagues Mahesh Anand, Dominic Fortes and Katherine Joy for helpful comments on an earlier draft of this manuscript. I thank Huma Irfan for help calculating areas on the lunar surface, Larry Nittler for help on meteorite compositions, Carsten Schwandt for advice on the electrochemical reduction of metal oxides, and colleagues who gave permission for me to reproduce images (acknowledged in the figure captions).